\RequirePackage{fix-cm}
\documentclass[smallextended]{svjour3}       
\smartqed  
\usepackage{graphicx}
\usepackage{graphicx}
\usepackage{amssymb}
\usepackage{graphicx}
\usepackage{subcaption}
\usepackage{multirow, hhline}
\usepackage{pifont}
\usepackage{hyperref}
\usepackage{listings}
\usepackage{authblk}

\let\vec\relax
\DeclareMathAccent{\vec}{\mathord}{letters}{"7E}
\makeatletter
\renewcommand{\subsection}{%
  \@startsection{subsection}
    {2}
    {\z@}
    {-21dd plus-8pt minus-4pt}
    {10.5dd}
    {\normalsize\bfseries\boldmath}%
}
\makeatother
\journalname{Journal}
\begin{document}

\title{Optimising Fine-Grained Access
Control Policy Enforcement for Database Queries.
A Model-Driven Approach}

\titlerunning{Optimising FGAC Policy Enforcement for Database Queries}        

\author{Hoang Phuoc-Bao Nguyen         \and
       Manuel Clavel 
}


\institute{Hoang Phuoc-Bao Nguyen \at
             Department of Computer Science \\
             ETH Zurich, Switzerland \\
              \email{hoang.nguyen@inf.ethz.ch}           
           \and
           Manuel Clavel \at
           Eastern International University \\
	\email{manuel.clavel@eiu.edu.vn}
}

\date{Received: date / Accepted: date}

\maketitle

\begin{abstract}
Recently, we have proposed 
 a model-driven approach for
 enforcing fine-grained access control (FGAC) policies 
 when executing SQL queries.
 More concretely,  we have defined 
  a function ${\rm SecQuery}()$ that,
  given an FGAC policy ${\cal S}$ and a SQL select-statement $q$,
  generates a SQL stored-procedure 
  $\ulcorner${\rm Sec\-Query}$({\cal S}, q)\urcorner$,
  such that:
  if a user $u$ with role $r$ is authorised, according to ${\cal S}$,
  to execute $q$ based on the current state of the database, 
  then calling  
  $\ulcorner{\rm SecQuery}({\cal S}, q)\urcorner(u,r)$
  returns the same result as when $u$ executes $q$;
  otherwise, if the user $u$ is not authorised, according to ${\cal S}$,
  to execute $q$ based on the current state of the database, 
  then calling 
  $\ulcorner{\rm SecQuery}({\cal S}, q)\urcorner(u,r)$
  signals an error.
  Not surprisingly, executing the query $q$ takes less time
  than calling the corresponding stored-procedure 
  $\ulcorner{\rm SecQuery}({\cal S}, q)\urcorner$. 
Here we propose a model-based methodology 
for \emph{optimising} the stored-procedures 
generated by the
function ${\rm SecQuery}()$. The idea is
to eliminate authorisation checks in the body of the
stored-procedures generated by ${\rm SecQuery}()$, when they
can be proved to be unnecessary. Based on our previous mapping
from the Object Constraint Language (OCL) to many-sorted first-order logic, we can attempt
to prove that authorisation checks are unnecessary by using
SMT solvers. We include a case study to illustrate and 
show the applicability of our methodology.
\end{abstract}

\section{Introduction}

Model-driven security (MDS)~\cite{BasinDL06,BasinCE11} specialises model-driven engineering for developing secure systems. In a nutshell, designers specify system models along with their security requirements and use tools to automatically generate security-related system artifacts, such as access control infrastructures. 


MDS has been applied with encouraging results to the 
development of \emph{data-centric applications}~\cite{BasinCEDD14}.
These applications are focused around  actions that create,
read, update, and delete data stored in a database. 
Data-centric applications are typically built following the so-called three-tier architecture.
 According to this architecture, applications consist of three layers: presentation layer, application layer, and data layer. 
The presentation layer helps to shape the look  of the application.
The application layer handles the application's business logic:  it defines the core functionality, 
and it acts as the middle layer connecting the presentation layer and the data layer. 
Lastly, the data layer is where information is stored through a database management system. 

When the data stored 
is sensitive, then the 
user's actions on these data  must be controlled.
If the access control policies are sufficiently simple, as in the case of role-based access control (RBAC)~\cite{Ferraiolo01} policies, 
it may be possible to formalise them declaratively, independent of the application's business logic. In contrast, \emph{fine-grained access control} (FGAC) policies may 
depend not only on the user's credentials but also on the satisfaction of constraints 
on the data stored in the database. In such cases, authorisation checks are often implemented programmatically, by  encoding them at appropriate places 
in the application layer. In our opinion, the following three reasons are recommended 
against this common practice. First of all, in order to perform the authorisation 
checks, the application layer must have full access (potentially) to the data stored
in the database. Secondly, in the case of FGAC policies, the application layer must perform the authorisation checks (potentially) for every row/cell, negatively impacting the overall performance of the application. Thirdly, changes in the access control policy will necessarily imply non-trivial changes in the application layer. 

In our opinion, a better approach for enforcing FGAC policies in data-centric applications is to perform the authorisation checks in the data layer for 
 the following reasons.~\footnote{About the importance of supporting  FGAC at the database level,
  we basically agree with~\cite{Kabra06}:
``Fine-grained access control [on databases] has traditionally been
performed at the level of application programs.  However, implementing
security at the application level makes management of authorization
quite difficult, in addition to presenting a large surface area for
attackers --- any breach of security at the application level exposes
the entire database to damage, since every part of the application has
complete access to the data belonging to every application
user.''} First of all, 
sensitive data will not be retrieved from the database in
an uncontrolled way  for the purpose of performing authorisation checks at the application layer. Secondly, FGAC checks will perform more efficiently at the database layer, levering on the  highly sophisticated optimisations for filtering data. Thirdly, changes in the access control policy will certainly imply changes in the data layer, but not in the application layer, which fits very well with the typical modularity and separation of concerns of a three-tier architecture.

Unfortunately, database-management systems do not currently provide  built-in 
features for enforcing FGAC policies. 
Broadly speaking, in the case of relational database-management systems,
the solutions currently offered are either (i) to manually create appropriate ``views'' in the database and to modify the queries to reference these views; or (ii) to use non-standard, proprietary enforcement mechanisms. 
These solutions are far from ideal. In fact, they are inefficient, error-prone, and scale poorly, as argued in~\cite{DBLP:journals/jot/BaoC20}.
%

We have  proposed in~\cite{DBLP:journals/jot/BaoC20} a \emph{model-driven approach for enforcing FGAC policies} when executing SQL queries.
In a nutshell,  we have defined  a function ${\rm SecQuery}()$ that,
given an FGAC policy ${\cal S}$ and a SQL select-statement $q$,
generates a SQL stored-procedure 
 $\ulcorner${\rm Sec\-Query}$({\cal S}, q)\urcorner$,
  such that:
  if a user $u$ with role $r$ is authorised, according to ${\cal S}$,
  to execute $q$ based on the current state of the database, then calling  
  $\ulcorner{\rm SecQuery}({\cal S}, q)\urcorner(u,r)$
  returns the same result as when $u$ executes $q$;
  otherwise, if the user $u$ is not authorised, according to ${\cal S}$,
  to execute $q$ based on the current state of the database, 
  then calling 
  $\ulcorner{\rm SecQuery}({\cal S}, q)\urcorner(u,r)$
  signals an error.
 The key features of our approach are the following: (i) The enforcement mechanism leaves unmodified the underlying database, except for adding the stored-procedures that configure the FGAC enforcement mechanism. (ii) The FGAC policies and the database queries are kept independent of each other, except that they refer to the same underlying data model. This means, in particular, that FGAC policies can be specified without knowing which database queries will be executed, and vice versa. (iii) The enforcement mechanism can be automatically generated from the FGAC policies.

There is, however, a  clear drawback in the approach proposed 
in~\cite{DBLP:journals/jot/BaoC20}.  
As mentioned before, FGAC policies depend not only on 
the assignments of users and permissions to roles, but also on 
the satisfaction of authorisation constraints on the current state of the database. 
Thus, a ``secured'' query generated by ${\rm SecQuery}()$ will typically include expressions in charge of checking that the relevant authorisation constraints 
are satisfied in the current state of the database. Unavoidably, executing these expressions will cause a performance penalty, greater or lesser depending on the 
``size'' of the database and the ``complexity'' of the corresponding authorisation constraints. There are, however, situations in which (some of) these authorisation checks seem unnecessary. 

In this article we propose a model-based methodology 
for \emph{optimizing} the stored-procedures 
generated by the
function ${\rm SecQuery}()$. The idea is
to eliminate authorisation checks from the body of the
stored-procedures generated by ${\rm SecQuery}()$, when they
can be \emph{proved} to be unnecessary, for which we propose to
use SMT solvers.
We report on a case study that illustrates  the applicability of our methodology.



  
  \subsubsection*{Organization}
  In Sections~\ref{fgac-models}--\ref{enforcing-fgac-authorisation} we 
 recall 
 our model-driven approach for
  enforcing FGAC policies when executing database 
  queries. 
  In particular: in Section~\ref{fgac-models} we introduce  FGAC security models;  
  in Section~\ref{fgac-authorization} we discuss FGAC authorisation for
  database queries; and  in Section~\ref{enforcing-fgac-authorisation} 
  we consider enforcing FGAC authorisation
  for database queries. The emphasis in 
  these sections
  is about the key
  components that conform to our model-driven approach,  
  and about their expected properties. 
  To illustrate and exemplify our approach, 
  we provide 
  concrete details of how 
  these  components are realised  in SQLSI --- a methodology 
  for enforcing FGAC policies when
  executing SQL queries.
 The interested reader can find the formal definitions of the SQLSI's key
 components in~\cite{DBLP:journals/jot/BaoC20}.
  
Then, in Section~\ref{optimising-fgac-authorisation} we present our  approach for optimising 
FGAC authorisation enforcement for database queries, and discuss  its 
realization in  SQLSI. 
%
Finally, in Section~\ref{case-study} we report  on a concrete case study showing
how our  approach can be applied for optimising SQLSI 
FGAC policies enforcement,
for different SQL queries and  FGAC policies. 
We conclude with related work and future work, 
in Sections~\ref{related-work} and~\ref{conclusions-future-work}.    
\section{FGAC security models}
\label{fgac-models}

A model-driven approach for enforcing FGAC policies for database queries
requires, in particular, that  FGAC policies are specified using 
\emph{models} and that the corresponding policy-enforcement artifacts 
are \emph{generated} from these models.

FGAC security models typically specify the \emph{resources}  to be protected, the \emph{scenarios} on which resources occur, the \emph{actions} on these resources to be controlled, 
and the \emph{authorisation constraints} to  control these actions.
%
%
FGAC security models also typically specify
the \emph{users} that can attempt to access  the resources, 
and the \emph{roles} that can be assigned to them.
%
In our general approach, we assume that authorisation constraints
are specified using \emph{expressions}, possibly containing keywords
denoting the resources being accessed and the user accessing it.
Moreover, we assume that there exists a  Boolean function
${\rm Eval}()$ such that, for any scenario ${\cal O}$,
any authorisation constraint ${\it auth}$, any user $u$, 
and any list of concrete resources $\vec{w}$, 
the function call 
${\rm Eval}({\cal O}, {\it auth}[u, \vec{w}])$ returns either true ($\top$)
or false ($\bot$), where ${\it auth}[u, \vec{w}]$ denotes the
expression ${\it auth}$ after substituting its keywords by the 
corresponding values in $u$, $\vec{w}$.


In our general approach, we  assume that each 
FGAC security model  defines a  Boolean function
${\rm Auth}()$ such that, for  any scenario ${\cal O}$,
 any user $u$, any role $r$,  any action ${a}$,
 and any list of concrete resource $\vec{w}$, 
the function call ${\rm Auth}({\cal O}, u, r, a, \vec{w})$
returns either true ($\top$) or false ($\bot$), indicating
whether the user $u$, with role $r$ is \emph{authorised} or not to 
perform the action ${a}$ on
the concrete resources $\vec{w}$ in the scenario ${\cal O}$.
Typically, the function ${\rm Auth}()$ will call the function
${\rm Eval}()$ for checking if the corresponding authorisation
constraint is satisfied or not.

  

\subsection*{FGAC security models in SQLSI}
 In SQLSI we use SecureUML~\cite{Lodderstedt02,BasinDL06} for modelling FGAC policies.
SecureUML is an extension of Role-Based Access Control (RBAC)~\cite{Ferraiolo01}. 
In RBAC, permissions are assigned to roles, and roles are assigned to users. In SecureUML, on the other hand,  one can model access control decisions that depend on two kinds of information: 
the assignments of users and permissions to roles; and 
the satisfaction of authorisation constraints by the current state of the database.

In SQLSI we model the resources to be protected using  \emph{data models},
 which consist of 
classes and associations, and we model scenarios as instances 
of these data models.
Currently, we do not support class generalisations, and 
we only consider  \emph{read}-actions on class attributes and association-ends as  actions to be controlled.

 
 \begin{figure}
\centering
\includegraphics[width=1\textwidth]{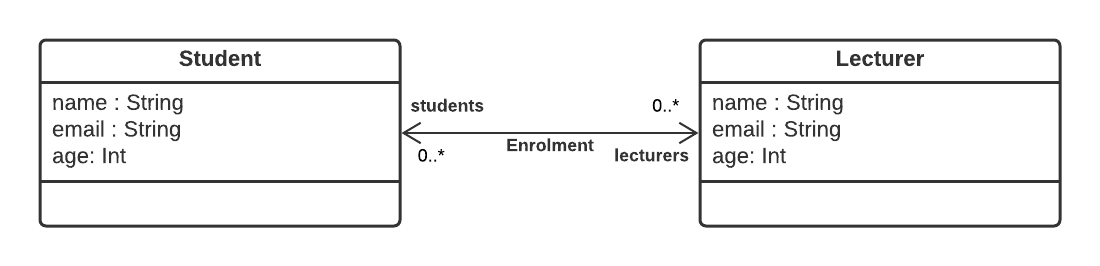}
 \caption{Example: the data model ${\tt University}$.}
 \label{university:dm}
 \end{figure}

\begin{example}
As a basic example, we introduce in Figure~\ref{university:dm} 
the data model {\tt Uni\-ver\-sity}.
 It contains two classes, 
{\tt Student} and {\tt Lecturer}, and one association 
{\tt Enrolment} between both of them. 
The classes {\tt Student} and {\tt Lecturer} have  attributes 
{\tt name}, {\tt email}, and {\tt age}. The class 
{\tt Student} represents the students of the university, with 
their name, email, and age. The class {\tt Lecturer} represents the lecturers of the university, with their name, email, and age. 
The association {\tt Enrolment} represents the relationship between the students (denoted by {\tt students}) and the lecturers (denoted by 
{\tt lecturers}) of the courses the students have enrolled in. 
\end{example}

In SQLSI we model authorisation constraints using the Object Constraint Language (OCL)~\cite{OCL14}. 
Authorisation constraints can contain keywords referring to 
resources --- namely, to the
object whose attribute is being accessed (denoted by the keyword
${\tt self}$), or to the objects linked by the association that is being accessed
(denoted by the corresponding association-ends). Authorisation constraint
can also contain keywords referring  to users
--- namely, to the user who is
attempting to access the resources (denoted by the keyword {\tt caller}). 
For the sake of clarity, in SQLSI we underline  keywords when they appear in 
authorisation constraints.

As expected, in SQLSI the  function ${\rm Eval}()$  corresponds to evaluating the given authorisation constraint in the given scenario
according to the standard semantics of OCL. 

\begin{example}
Consider the following security model 
{\tt SecVGU\#A} for the data model ${\tt University}$.
\begin{itemize}
\item Roles. There is only one role, namely, the role ${\tt Lecturer}$. Lecturers are assigned to this role.
\item Permission:
\begin{itemize}
\item \emph{Any lecturer can know his/her students}. More formally, for a user 
{\tt caller} with role {\tt Lecturer} to read the resources linked by the association 
{\tt Enrolment}, the following OCL constraint 
must be satisfied:

\begin{tabbing}
$\underline{{\tt lecturers}}$ ${\tt =}$ $\underline{{\tt caller}}$,
\end{tabbing}

in which, as explained before, $\underline{{\tt lecturers}}$ is
a keyword denoting any lecturer linked by the association {\tt Enrolment} at its association-end {\tt lecturers}.

\item \emph{Any lecturer can know his/her own email, as well as the emails of his/her students}. More formally, for a user  {\tt caller} with role
{\tt Lecturer} to read the email of a lecturer's resource {\tt self}, the
following OCL constraint must be satisfied:

\begin{tabbing}
$\underline{{\tt caller}}$ ${\tt =}$ $\underline{{\tt self}}$,
\end{tabbing}

\item \emph{Any lecturer can know the emails of his/her students}. More formally, for a user {\tt caller} with role
{\tt Lecturer} to read the email of a student's resource {\tt self}, the
following OCL constraint must be satisfied:

\begin{tabbing}
$\underline{{\tt caller}}{\tt .students}\rightarrow{\tt  includes(\underline{{\tt self}})}$.
\end{tabbing}
\end{itemize}
\end{itemize}
\end{example}

\section{FGAC-authorisation for database queries}
\label{fgac-authorization}

As expected, in our general approach we assume that 
databases are used for storing information, and
that they provide different means to manage this information.
In particular, we assume that they support \emph{queries}  for information retrieval. 
More specifically, we assume that there exists a   function
${\rm Exec}()$
such that given a database instance ${\it db}$
and a query $q$, the
function call ${\rm Exec}({\it db},  q)$  either returns the \emph{result} of executing 
the query $q$ in the database instance ${\it db}$, or  it returns an \emph{error}.

In our general approach, we assume that there exists 
a Boolean function ${\rm AuthQuery}()$  such that, given an FGAC security model
${\cal S}$, a  query $q$,  a database instance ${\it db}$, 
a user $u$,  and a role $r$, 
the function call ${\rm AuthQuery}({\cal S}, u, r,$ $q, {\it db})$ 
returns either true ($\top$) or false ($\bot$), indicating whether the user $u$, 
with role $r$ is authorised or not to execute the query $q$ 
in the database instance ${\it db}$.

\subsection*{FGAC-authorisation for database queries in SQLSI}

In SQLSI we consider SQL queries.
The SQLSI's definition of the function ${\rm AuthQuery}()$~\cite{DBLP:journals/jot/BaoC20} is based on the following 
consideration.
A user can be authorised to execute a
query on a database if \emph{the execution of this query does not leak
confidential information}, according to the given FGAC policy.
However, this typically implies 
much more than simply
checking that the final result satisfies the given FGAC policy,
since a clever attacker can devise a query such that the simple
fact that a final result is obtained
may reveal some confidential information.
To illustrate this point, 
consider the select-statements in Figures~\ref{query1:fig}--\ref{query3:fig}.~\footnote{For the sake of readability, we have
formalised  these queries using the  \emph{names}
of the students and the lecturers,  instead of their 
database \emph{ids}.}
Suppose that, for a given scenario, 
the three select-statements return the same final result, namely, a non-empty string,
 representing an email, which is not considered confidential. On a closer examination,
  however, we can realise that, for each of these select-statements, 
  the final result is revealing additional information, 
which may in turn be confidential. In particular,

\begin{itemize}
\item {\tt Query\#1} reveals that the returned email belongs to Huong.
\item {\tt Query\#2} reveals not only that the returned email belongs to Huong, but also that Thanh is enrolled in a course that Huong is teaching. 
\item {\tt Query\#3} reveals that the email belongs to Huong, and that Huong and Manuel are ``colleagues'', in the sense that there are some students who have both Huong and Manuel as their lecturers.
\end{itemize}

\begin{figure}
\begin{subfigure}{\linewidth}
\begin{lstlisting}[numbers=none]
SELECT email FROM Lecturer WHERE Lecturer_id = 'Huong'
\end{lstlisting}
\caption{The query {\tt Query\#1}.}
\label{query1:fig}
\end{subfigure}\\[3ex]
\begin{subfigure}{\linewidth}
\begin{lstlisting}[numbers=none]
SELECT DISTINCT email FROM Lecturer
JOIN (SELECT * FROM Enrolment
      WHERE students = 'Thanh' AND lecturers = 'Huong') AS TEMP
ON TEMP.lecturers = Lecturer_id
\end{lstlisting}
\caption{The query {\tt Query\#2}.}
\label{query2:fig}
\end{subfigure}\\[3ex]
\begin{subfigure}{\linewidth}
\begin{lstlisting}[numbers=none]
SELECT DISTINCT email FROM Lecturer
JOIN (SELECT huong_enrolments.lecturers AS lecturers
      FROM (SELECT * FROM Enrolment
            WHERE lecturers = 'Manuel') AS manuel_enrolments
      JOIN (SELECT * FROM Enrolment
            WHERE lecturers = 'Huong') AS huong_enrolments
      ON manuel_enrolments.students = huong_enrolments.students
) AS TEMP
ON TEMP.lecturers = Lecturer_id
\end{lstlisting}
\caption{The query {\tt Query\#3}.}
\label{query3:fig}
\end{subfigure}
\caption{Example queries}
\end{figure}

%
In fact, the SQLSI's function ${\rm AuthQuery}()$ is defined in such a way
that any information that may be used to reach the final result of a query 
(in particular, any information involved in subqueries, 
where-clauses, and on-clauses) 
is  checked for policy-compliance.
In this way, for example, if a user is not authorised to know whether Huong is Thanh's lecturer or not, then he/she will not be  authorised to execute 
{\tt Query\#2}, even when he/she may be authorised
to access Huong's email. 
Similarly, if a user is not authorised to know whether Huong and Manuel are ``colleagues'' or not, then, he/she will not be authorised,
to execute 
{\tt Query\#3}, even when he/she may be authorised to access lecturers' emails.~\footnote{The SQLSI's 
function ${\rm AuthQuery}()$ does
not preclude the possibility that,
if an attacker \emph{knows the specific
{FGAC} policy being enforced}, he/she can 
devise a query such that a ``non-authorised'' response may
still leak confidential information. }

\section{Enforcing FGAC-authorisation for database queries}
\label{enforcing-fgac-authorisation}

In our general approach, the FGAC enforcement mechanism
for database queries consists of generating  ``secured'' versions of the 
given queries, 
and then executing these ``secured'' versions 
instead of the given queries.
More specifically, we consider the following notion of ``secured'' queries.
Given an FGAC model ${\cal S}$, a database query $q$, and a 
database instance ${\it db}$,
we say that $q^{\flat}$ is a \emph{secured} version 
of a query $q$, if and only if, for any user $u$, and any
role $r$:
\begin{itemize}
\item if ${\rm AuthQuery}({\cal S}, u, r, q, {\it db}) = \bot$, 
then ${\rm Exec}({\it db}, q^{\flat})$ returns an error.
\item otherwise, ${\rm Exec}({\it db},  q^{\flat}) = {\rm Exec}({\it db}, q)$.
\end{itemize}

In our general approach, we assume that there exists 
a function ${\rm SecQuery}()$ such that, given an FGAC security model
${\cal S}$ and a database query $q$,  the function call
 ${\rm SecQuery}({\cal S}, q)$ 
 returns a ``secured'' version of the query $q$.
%
 
 \subsection*{Enforcing FGAC-authorisation for SQL queries in SQLSI}
%
In SQLSI, given an FGAC security model $S$ and a SQL query $q$,
the function ${\rm SecQuery}()$~\cite{DBLP:journals/jot/BaoC20}
generates a SQL stored-procedure 
 $\ulcorner{\rm Sec\-Query}({\cal S}, q)\urcorner$
 that implements the \emph{authorisation checks}
   required by the SQLSI's function
  ${\rm AuthQuery}()$ to comply with policy ${\cal S}$
  when executing the query $q$.

\section{Optimising FGAC policy  enforcement for data\-base queries}
\label{optimising-fgac-authorisation}
As explained before, FGAC policies depend not only on 
the assignments of users and permissions to roles, but also on 
the satisfaction of authorisation constraints on the current state of the system. 
Therefore, in our general approach, we assume that the ``secured'' 
queries generated by ${\rm SecQuery}()$   include 
expressions in charge of checking that the
relevant authorisation constraints are satisfied in the current
state of the database.
More specifically, we assume, first of all, 
that there exists a one-to-one correspondence 
between the  data model's scenarios  and the database instances.
%
We also assume that there exists a one-to-one correspondence 
between the users and roles declared in the FGAC security model
and those declared in the database. 
Then, we assume that
there is a function ${\rm map}()$ such
that, for any authorisation constraint ${\it auth}$,
${\rm map}({\it auth})$
returns a \emph{correct implementation} of ${\it auth}$,
in the following sense:
for any scenario ${\cal O}$, any user $u$,  any concrete resources $\vec{w}$,
and any assignment $\sigma = \{{\tt caller}\mapsto u, \vec{k} \mapsto \vec{w}\}$,
\begin{eqnarray}
\label{correct-implementation}
{\rm Eval}({\cal O}, \sigma({\it auth}))
= \top
\Longleftrightarrow
{\rm Exec}_{\overline{\sigma}} (\overline{{\cal O}}, {\rm map}({\it auth}))
= {\tt TRUE}
\end{eqnarray}
where $\overline{{\cal O}}$ denotes the database instance corresponding
to the scenario ${\cal O}$, and
${\rm Exec}_{\overline{\sigma}}$ denotes the execution-context
where the keywords  in  ${\it auth}$ 
are
assigned  values according to the assignment $\sigma$.
%



Consider now the \emph{cost} of executing the ``secured'' queries 
generated by ${\rm SecQuery}()$. 
As mentioned before, these queries include expressions generated by
${\rm map}()$ for checking that the relevant authorisation
constraints  are satisfied in the current state of the database.
Unavoidably,  these expressions cause a performance penalty at execution-time,
greater or lesser depending on the ``size'' of the database 
and on their own ``complexity''.
There are, however, situations in which these \emph{expensive} authorisation checks seem unnecessary.
%
%
%
Notice, in particular that, for any authorisation constraint
${\it auth}$, 
we can safely eliminate the authorisation check ${\rm map}({\it auth})$
--- based on the correctness assumption~(\ref{correct-implementation}) ---,
 if we can  prove that, for any scenario ${\cal O}$, any user $u$, any concrete resources $\vec{w}$, it holds that
${\rm Eval}({\cal O}, \sigma({\it auth}))
= \top$.
%
Interestingly, 
${\rm Eval}({\cal O}, \sigma({\it auth})) = \top$ may only hold for
scenarios ${\cal O}$ which satisfies certain known properties:
for example, that every student is over 21 years old.
In these cases, the elimination of the authorisation check ${\rm map}({\it auth})$
is only safe 
if the aforementioned properties can be guaranteed to be satisfied by the database 
  when the query is executed.
Similarly, ${\rm Eval}({\cal O}, \sigma({\it auth}))
= \top$ may only hold for
users $u$ and/or resources $\vec{w}$
 which satisfies certain known properties:
for example, that the lecturer attempting to execute the
 query is the oldest lecturer in the university, or 
that the query is only about  students enrolled in some classes  
of the lecturer attempting to execute the query.
As before, the elimination of the authorisation check ${\rm map}({\it auth})$
is only  safe 
if the aforementioned properties can be guaranteed to be satisfied by the database  
when the query is executed.

\subsection*{Optimising FGAC policy  enforcement for data\-base queries in
SQLSI}

The SQLSI's  mapping from data models
to SQL schemas is defined in~\cite{DBLP:journals/jot/BaoC20}. In a nutshell,
classes are mapped to tables, attributes to columns,
and many-to-many associations to tables with
the corresponding foreign-keys, in such a way that
objects and links can be stored, respectively,
 in the tables corresponding to their classes and 
 the tables corresponding to their associations.
Tables corresponding to classes 
contain an extra column to store the objects' unique identifiers.
The name of this extra column is the table's name followed by ${\tt \_id}$.
%

As for  the function ${\rm map}()$, in charge of implementing in SQL
the  OCL authorisation constraints, we  can 
\emph{reuse},  of course, the available mappings
from OCL to SQL --- for example~\cite{BaoC19}.
However, for the sake of execution-time performance, 
we recommend manually implementing in SQL the OCL authorisation
constraints, and to take responsibility for its correctness.

 
 
Finally, we propose  to use the mappings from OCL to many-sorted first-order logic (MSFOL) introduced in~\cite{DaniaC16} for proving 
that authorisation checks are unnecessary in the ``secured'' queries generated
 by ${\rm SecQuery}()$, and therefore can be safely removed.
In a nutshell,~\cite{DaniaC16} defines the following mappings:
(i) a mapping ${\rm map}()$ from data models  to  MSFOL theories;
(ii) a mapping ${\rm intr}()$ from 
scenarios  to  MSFOL interpretations;
and   (iii) a mapping ${\rm map}_{{\rm true}}()$ 
from OCL boolean expressions 
to MSFOL formulas.
In the case of an expression ${\it exp}$ containing
collection sub-expressions, the formula
 ${\rm map}_{{\rm true}}({\it exp})$ will
contain the corresponding  predicate expressions;
the conjunction of formulas defining 
these predicates is generated by
a mapping ${\rm map}_{\rm def}()$ which is defined along with
the mapping ${\rm map}_{{\rm true}}()$.

%
The mappings introduced in~\cite{DaniaC16} 
satisfy the following property:
let ${\cal D}$ be a data model, and let ${\cal O}$ be a scenario
 of ${\cal D}$. Let ${\it exp}$ be a ground (i.e., no free variables)
 boolean OCL expression. Then, the following holds:
\begin{eqnarray}
\label{correctness-map-true}
{\rm intr}({\cal O})\models 
({\rm map}_{{\rm def}}({\rm exp})
\Rightarrow
{\rm map}_{{\rm true}}({\rm exp}))
\Longleftrightarrow
{\rm Eval}({\cal O}, {\it exp}) = \top.
\end{eqnarray}
\noindent Hence, when deciding whether the authorisation check corresponding
to an authorisation constraint ${\it auth}$ is unnecessary and therefore can be
safely removed from the ``secured'' queries generated from the ${\rm SecQuery}()$,
we can reduce the problem of proving that for
any scenario ${\cal O}$, any user $u$,
any concrete resources $\vec{w}$, 
and any assignment $\sigma = \{{\tt caller}\mapsto u, \vec{k} \mapsto \vec{w}\})$ holds that: 
\begin{eqnarray*}
{\rm Eval}({\cal O}, \sigma({\it auth}))
= \top,
\end{eqnarray*}
to the problem of proving that the following MSFOL theory is \emph{unsatisfiable}:
\begin{eqnarray}
\label{unnecessary:sat}
{\rm map}({\cal D}, \sigma) 
\wedge
{\rm map}_{{\rm def}}({\it auth}) \wedge
\neg({\rm map}_{{\rm true}}({\it auth})),
\end{eqnarray}
\noindent where ${\rm map}({\cal D}, \sigma)$
simply adds to the MSFOL theory ${\rm map}({\cal D})$ 
the constant symbols ${\tt caller}$ and $\vec{k}$, 
with the appropriate sort declarations.
Then, if~(\ref{unnecessary:sat}) is unsatisfiable, 
we can safely conclude that the authorisation check
corresponding to the constraint ${\it auth}$ is indeed
unnecessary, since ${\it auth}$ cannot be false in any scenario.

In the following section we   present a case study in which we
apply the above methodology
to safely eliminate unnecessary authorisation checks
from  ``secured'' queries generated by the SQLSI's function 
${\rm SecQuery}()$.
Interestingly,  
the authorisation checks that we consider in our case study 
seem to be  unnecessary only for  scenarios,  users, or 
 resources that satisfy certain known properties.
As expected, to prove that they are indeed unnecessary
in these cases
we formalise the known properties 
as OCL boolean expressions,
map these expressions into MSFOL formulas,
and  join (with a conjunction) these formulas to the corresponding
satisfiability problem.


\section{Case study}
\label{case-study}

In this section we apply to 
different FGAC policies, different users, and  different queries  
the methodology introduced above
for optimising   ``secured'' queries generated 
by the SQLSI's function  ${\rm SecQuery}()$.

We first introduce  
two different  policies
for the data model 
\texttt{University} shown in~Figure~\ref{university:dm}.
%
\begin{itemize}
%

\item The policy  $\texttt{SecVGU\#1}$
contains the following 
clauses: (i) \emph{a lecturer can know the age of any student, 
if no other lecturer is older than  he/she is}; and
(ii) \emph{a lecturer can know the students of any lecturer, 
if no other lecturer is older than he/she is}.
This policy can be modelled in SQLSI as follows:

\begin{tabbing}
 ${\rm roles}$  = $\{ {\tt Lecturer}\}$\\
 ${\rm auth}({\tt Lecturer}, {\rm read}({\tt Enrolment}))$\\
 \quad= ${\tt Lecturer.allInstances()}
 {\tt \rightarrow{}select(l\mid l.age>{\mbox{\underline{\tt caller}}}.age)}$\\
 \qquad${\tt \rightarrow{}isEmpty()}$\\
${\rm auth}({\tt Lecturer}, {\rm read}({\tt Student:age}))$\\
 \quad= ${\tt Lecturer.allInstances()}
 {\tt \rightarrow{}select(l\mid l.age>{\mbox{\underline{\tt caller}}}.age)}$\\
 \qquad ${\tt \rightarrow{}isEmpty()}$
\end{tabbing}

\item the policy, $\texttt{SecVGU\#2}$
 contains the following clauses:
(i) \emph{a lecturer can know the age of any student, if the student is his/her student}; 
(ii) \emph{a lecturer can know his/her students}; and (iii) \emph{a lecturer can know the students of any lecturer if the student is his/her student}.
This policy can be modelled in SQLSI as follows:
  
  \begin{tabbing}
  ${\rm roles}$  = $\{ {\tt Lecturer}\}$\\
  ${\rm auth}({\tt Lecturer}, {\rm read}({\tt Student\!:\!age}))$\\
$ \quad =  \mbox{\underline{\tt caller}}
 {\tt .students\rightarrow{}{\tt exists}(s\mid{}s = \mbox{\underline{\tt self}})}$\\
  ${\rm auth}({\tt Lecturer}, {\rm read}({\tt Enrolment}))$\\
$\quad = {\tt (\mbox{\underline{\tt caller}} = \mbox{\underline{\tt lecturers}}})\ {\tt or}$\\
$\qquad {\tt (\mbox{\underline{\tt caller}}.students\rightarrow{}{\tt exists}(s\mid{}s = \mbox{\underline{\tt students}}))}$
   \end{tabbing}
   
 \end{itemize}

  Next we introduce three different SQL queries for the 
  database corresponding to the data model ${\tt University}$.
  
  \begin{itemize}
  \item the query {\tt Query\#4} that asks
  \emph{the number of students
  whose age is greater than 18}. This query can be
  expressed in SQL as follows:
  

\begin{lstlisting}[numbers=none]
SELECT COUNT(*) FROM Student WHERE age > 18
\end{lstlisting}
  
  \item the query {\tt Query\#5} that asks
  \emph{the number of enrolments}. This query can be
  expressed in SQL as follows:

  
\begin{lstlisting}[numbers=none]
SELECT COUNT(*) FROM Enrolment
\end{lstlisting}
  
  \item the query {\tt Query\#6} that asks
  \emph{the age of the students of the user assigned to
  the variable {\tt caller}}. This query can be
  expressed in SQL as follows:

\begin{lstlisting}[numbers=none]
SELECT age FROM Student
JOIN (SELECT * FROM Enrolment
      WHERE lecturers = caller) AS my_enrolments
ON my_enrolments.students = Student_id
\end{lstlisting}
  
  \end{itemize}

Finally, in order to follow the  case study, 
we  recall  here the main
``features'' of  the stored-procedures generated by
the SQLSI's function ${\rm SecQuery}()$.
The interested readers can find the full definition
of the SQLSI's function ${\rm SecQuery}()$ in~\cite{DBLP:journals/jot/BaoC20}.
A stored-procedure
  generated by  ${\rm SecQuery}()$
has two parameters  {\tt caller} and {\tt role}, which represent,
 respectively, the user executing the given query and
 the role of this user when executing this query.
The  body of the stored-procedure 
creates a list of temporary tables and, if successful, it
executes  the original query.
These temporary tables 
correspond to the conditions
that need to be satisfied for the user, with the given role,
to be authorised to execute the given query.
The definition of each temporary table is such that,
when attempting to create the table, if the corresponding
 condition is not satisfied, then an error is signalled.
%
The reason
for using temporary tables instead of subqueries
is to prevent the SQL optimiser from ``skipping''
the authorisation checks that ${\rm SecQuery}()$  generates.
%
These authorisation checks are
 implemented
using case-expressions. 
Each of these case-expression calls a function ${\rm AuthFunc}()$,
which implements the authorisation constraint controlling the
access to the corresponding resource (attribute or association).
 If the result of this function call is \texttt{TRUE}, 
 then the case-expression  returns the
 requested resource; otherwise, it signals an error.
 As expected, for each authorisation constraint ${\it auth}$,
 the  function ${\rm AuthFunc}()$ executes ${\rm map}({\it auth})$, i.e.,
 the provided implementation in SQL of the OCL constraint ${\it auth}$.

\subsection{Case~1: {\tt Query\#4}} \label{cs1}
Let ${\cal S}$ be an  FGAC security model. We show below
the stored-procedure generated by the SQLSI's function 
${\rm SecQuery}()$ for {\tt Query\#4}.


\begin{lstlisting}[numbers=left, escapeinside={[*}{*]}]
CREATE PROCEDURE [*$\ulcorner{\rm SecQuery}({\cal S}, \texttt{Query\#4})\urcorner$*]
  (in caller varchar(250), in role varchar(250))
BEGIN
  DECLARE _rollback int DEFAULT 0;
  DECLARE EXIT HANDLER FOR SQLEXCEPTION
  BEGIN
    SET _rollback = 1;
    GET STACKED DIAGNOSTICS CONDITION 1
      @p1 = RETURNED_SQLSTATE, @p2 = MESSAGE_TEXT;
    SELECT @p1, @p2;
    ROLLBACK;
  END;
  START TRANSACTION;

  CREATE TEMPORARY TABLE TEMP1 AS (
    SELECT * FROM Student
    WHERE CASE [*$\ulcorner{\rm AuthFunc}({\cal S}, {\tt Student:age})\urcorner$*](caller, role, 
      Student_id) WHEN 1 THEN age ELSE throw_error() END > 18
  );

  CREATE TEMPORARY TABLE TEMP2 AS (
    SELECT Student_id AS Student_id FROM TEMP1
  );

  IF _rollback = 0
    THEN SELECT COUNT(*) from TEMP2;
  END IF;
END
\end{lstlisting}


%
%
%
\noindent Notice that,
when creating the temporary table {\tt TEMP1} in lines 15--19,
the SQL function $\ulcorner{\rm AuthFunc}({\cal S}, {\tt Student:age})\urcorner$
is called  for each row contained in the table
{\tt Student}. 
Therefore, 
the execution-time for
$\ulcorner{\rm SecQuery}({\cal S}, \texttt{Query\#4})\urcorner$ 
will increase
depending on the ``size'' of the table {\tt Student}
and   the ``complexity'' of the
SQL expression
${\rm map}({\rm auth}({\cal S}, r, {\rm read}({\tt Student\!:\!age})))$.

Consider the case of the policy {\tt SecVGU\#1}.
Recall that the authorisation constraint ${\rm auth}(\texttt{SecVGU\#1}, {\tt Lecturer}, 
{\rm read}({\tt Student\!:\!age}))$ is specified in OCL as follows:

\begin{tabbing}
 ${\tt Lecturer.allInstances()}
 {\tt \rightarrow{}select(l\mid l.age>{\mbox{\underline{\tt caller}}}.age)\rightarrow{}isEmpty()}$
\end{tabbing}

\noindent Suppose that we implement this constraint in SQL as follows:
%

\begin{lstlisting}[numbers=none, label = sqlqvinh5]
(SELECT MAX(age) FROM Lecturer)
   = (SELECT age FROM Lecturer WHERE Lecturer_id = caller)
\end{lstlisting}

%
\noindent Notice then that, when executing
\begin{eqnarray*}
\label{qvinh5}
\begin{array}{l}
{\ulcorner {\rm SecQuery}(\texttt{SecVGU\#1}, \texttt{Query\#4})
\urcorner}({\tt caller}, \mbox{``{\tt Lecturer}''}),
\end{array}
\end{eqnarray*}
%
the SQL expression above
will be executed for
each row in the table ${\tt Student}$. Moreover, notice that each time
this 
expression
is executed, 
the clause
\begin{lstlisting}[numbers=none]
WHERE Lecturer_id = caller
\end{lstlisting}
will make a search among the rows in the table {\tt Lecturer}.


\paragraph{Possible optimisations.}
Suppose that the user attempting to execute 
{\tt Query\#4} is 
the \emph{oldest lecturer}. 
In this case, 
the case-statement in lines 17--18 seems unnecessary, because
the policy {\tt SecVGU\#1} authorises a lecturer to know the age of every student, \emph{if no other lecturer is older than he/she is}.

Applying the methodology described above, and 
adding to the corresponding satisfiability problem
the fact that the user  is the oldest lecturer,
 we  can prove that  the case-statement in lines 17--18 
is indeed unnecessary, and therefore can be safely removed,
if the user attempting to execute the query is the oldest lecturer. 
%
The SMT solver CVC4~\cite{DBLP:conf/cav/BarrettCDHJKRT11} solves this
problem in 0.163 seconds.  
The interested reader can find in Listing~\ref{app:cs1:msfol} 
(Appendix~\ref{satisfiability:appendix}) 
the input to the CVC4 tool, and in Listing~\ref{app:cs1:optsp} 
(Appendix~\ref{optimised:appendix}) 
the optimised stored-procedure 
${\ulcorner{\rm SecQuery}(\texttt{SecVGU\#1}, \texttt{Query\#4})
\urcorner}$.

Finally, notice that the case-statement in lines 17--18 cannot be removed, 
however, for the case of the policy {\tt SecVGU\#2}, even if 
the user who is attempting to execute the query {\tt Query\#4} is 
the oldest lecturer.
The interested reader can find  in Listing~\ref{app:cs1:msfol:x3} 
(Appendix~\ref{satisfiability:appendix}) the satisfiability problem that corresponds to this case.

\subsection{Case~2: {\tt Query\#5}} \label{cs2}
Let ${\cal S}$ be an FGAC security model.
We show below the stored-procedure generated by the SQLSI's function ${\rm SecQuery}()$ for {\tt Query\#5}.

\begin{lstlisting}[numbers=left, escapeinside={[*}{*]}]
CREATE PROCEDURE [*$\ulcorner{\rm SecQuery}({\cal S}, \texttt{Query\#5})\urcorner$*]
  (in caller varchar(250), in role varchar(250))
BEGIN
DECLARE _rollback int DEFAULT 0;
DECLARE EXIT HANDLER FOR SQLEXCEPTION
BEGIN
  SET _rollback = 1;
  GET STACKED DIAGNOSTICS CONDITION 1 
    @p1 = RETURNED_SQLSTATE, @p2 = MESSAGE_TEXT;
  SELECT @p1, @p2;
  ROLLBACK;
  END;
  START TRANSACTION;

  CREATE TEMPORARY TABLE TEMP1 AS (
    SELECT Lecturer_id AS lecturers, Student_id AS students 
    FROM Lecturer, Student
  );

  CREATE TEMPORARY TABLE TEMP2 AS (
    SELECT * FROM TEMP1
    WHERE CASE [*$\ulcorner{\rm AuthFunc}({\cal S},$ ${\tt Enrolment})\urcorner$*](caller, role, 
      lecturers, students) WHEN TRUE THEN TRUE 
      ELSE throw_error() END
  );

  CREATE TEMPORARY TABLE TEMP3 AS (
    SELECT students FROM Enrolment
  );

  IF _rollback = 0
    THEN SELECT COUNT(*) from TEMP3;
  END IF;
END
\end{lstlisting}
%
%
%
%
%

%
Notice that,
when creating the temporary table {\tt TEMP2},
the function call $\ulcorner{\rm AuthFunc}({\cal S},$ ${\tt Enrolment})\urcorner$ 
is executed once for each record contained in the  table
${\tt TEMP1}$, which is defined as the cartesian product of the
tables ${\tt Student}$ and ${\tt Lecturer}$.
Therefore, the execution-time for
$\ulcorner{\rm SecQuery}({\cal S},$ $\texttt{Query\#5})\urcorner$ 
will increase
depending on the ``size'' of the tables ${\tt Student}$ and ${\tt Lecturer}$,
and the ``complexity'' of the SQL expression 
${\rm map}({\rm auth}({\cal S}, r, {\rm read}({\tt Enrolment})))$.

Consider the case of the policy {\tt SecVGU\#2}.
Recall that the authorisation constraint ${\rm auth}(\texttt{SecVGU\#2}, {\tt Lecturer}, 
{\rm read}({\tt Enrolment}))$ is specified in OCL as follows:
\begin{tabbing}
${\tt (\mbox{\underline{\tt caller}} = \mbox{\underline{\tt lecturers}})\ or\ (\mbox{\underline{\tt caller}}.students\rightarrow{}{\tt exists}(s\mid{}s = \mbox{\underline{\tt students}}))}$.
 \end{tabbing}
 
 Suppose that
we implement this authorisation constraint in SQL as follows:


\begin{lstlisting}[numbers=none]
(caller = lecturers)
OR (EXISTS (SELECT 1 FROM Enrolment e
           WHERE e.lecturers = caller
           AND e.students = students))
\end{lstlisting}
%
%
\noindent Notice then that, when executing
\begin{eqnarray*}
\begin{array}{l}
{\ulcorner {\rm SecQuery}(\texttt{SecVGU\#2}, \texttt{Query\#5})
\urcorner}({\tt caller}, \mbox{``{\tt Lecturer}''}),
\end{array}
\end{eqnarray*}
%
the SQL expression above will be executed once for
each row in the table {\tt TEMP1}, and that each time
this expression is executed, the clause 
\begin{lstlisting}[numbers=none]
WHERE e.lecturers = caller
AND e.students = students
\end{lstlisting}
will make a search among the rows in the table {\tt Enrolment}.

\paragraph{Possible optimisations}
Suppose now that the user who is attempting to execute the
query {\tt Query\#5} is \emph{a lecturer of every student}.
In this case 
the case-statement in lines 22--24 seems unnecessary, 
because the policy {\tt SecVGU\#2} authorises every lecturer to know 
the students of any lecturer if they are his/her students.

Applying the methodology described above,
adding to the satisfiability problem the fact that the
user who is attempting to execute the query is a lecturer of every
student,
we can in fact prove that the case-statement in lines 22--24 is indeed unnecessary,
and therefore can be safely removed if the user attempting to execute the 
query is a lecturer of every student.
The SMT solver CVC4~\cite{DBLP:conf/cav/BarrettCDHJKRT11} solves this satisfiability problem 
in $0.046$ seconds.
The interested reader can find in Listing~\ref{app:cs2:msfol} 
(Appendix~\ref{satisfiability:appendix})
the input to the CVC4 tool, and in Listing~\ref{app:cs2:optsp} 
(Appendix~\ref{optimised:appendix})
the optimised stored-procedure ${\ulcorner {\rm SecQuery}(\texttt{SecVGU\#2}, \texttt{Query\#5})\urcorner}$.
%

%
Finally, notice that the case-statement in lines 22--24 cannot be removed, however, 
 for the case of the policy {\tt SecVGU\#1}, even if the user who is attempting to execute the query {\tt Query\#5} is a lecturer of every student.
 The interested reader can find in 
 Listing~\ref{app:cs2:msfol:x2} (Appendix~\ref{satisfiability:appendix})
 the 
satisfiability problem that corresponds to this case.

\subsection{Case~3: {\tt Query\#6}} \label{cs3}

Let ${\cal S}$ be an FGAC security model. We show below the stored-procedure generated by the SQLSI's function ${\rm SecQuery}()$ for {\tt Query\#6}.

\begin{lstlisting}[numbers=left, escapeinside={[*}{*]}]
CREATE PROCEDURE [*$\ulcorner{\rm SecQuery}({\cal S}, \texttt{Query\#6})\urcorner$*] 
  (in caller varchar(250), in role varchar(250))
BEGIN
DECLARE _rollback int DEFAULT 0;
DECLARE EXIT HANDLER FOR SQLEXCEPTION
BEGIN
  SET _rollback = 1;
  GET STACKED DIAGNOSTICS CONDITION 1 
    @p1 = RETURNED_SQLSTATE, @p2 = MESSAGE_TEXT;
  SELECT @p1, @p2;
  ROLLBACK;
  END;
  START TRANSACTION;

  CREATE TEMPORARY TABLE TEMP1 AS (
    SELECT Student_id AS students, Lecturer_id AS lecturers
    FROM Student, Lecturer
    WHERE Lecturer_id = caller
  );

  CREATE TEMPORARY TABLE TEMP2 AS (
    SELECT * FROM TEMP1
    WHERE CASE [*$\ulcorner{\rm AuthFunc}({\cal S},$ ${\tt Enrolment})\urcorner$*](caller, role, 
      lecturers, students) WHEN TRUE THEN TRUE 
      ELSE throw_error() END
  );

  CREATE TEMPORARY TABLE TEMP3 AS (
    SELECT * FROM Enrolment WHERE lecturers = caller
  );

  CREATE TEMPORARY TABLE TEMP4 AS (
    SELECT * FROM Student JOIN TEMP3 
    ON Student_id = students
  );

  CREATE TEMPORARY TABLE TEMP5 AS (
    SELECT CASE [*$\ulcorner{\rm AuthFunc}({\cal S},$ ${\tt Student:age})\urcorner$*](caller, role,
      Student_id) WHEN 1 THEN age ELSE throw_error() END as age 
    FROM TEMP4
  );

  IF _rollback = 0
    THEN SELECT age from TEMP5;
  END IF;
END
\end{lstlisting}

Notice that, when creating the temporary table {\tt TEMP2}, 
the function call 
$\ulcorner{\rm AuthFunc}({\cal S}, {\tt Enrolment})\urcorner$
 is executed once for each record contained in the table 
 ${\tt TEMP1}$, which is defined as the subset of the 
  cartesian product of the tables ${\tt Student}$ and ${\tt Lecturer}$
  that contains only the students of the lecturer attempting to
  execute the query. Therefore, the execution-time for the stored-procedure
$\ulcorner{\rm SecQuery}({\cal S}, \texttt{Query\#6})\urcorner$ will increase depending 
on the ``size'' of the table ${\tt Student}$ and the 
``complexity'' of the implemented SQL expression ${\rm map}({\rm auth}({\cal S}, r, {\rm read}({\tt Enrolment})))$.

Similarly, notice that, when creating the temporary table 
${\tt TEMP5}$, the function call 
$\ulcorner{\rm AuthFunc}({\cal S}, {\tt Student:age})\urcorner$
is executed once for each record contained in the table 
${\tt TEMP4}$, which is defined as the join of the tables 
${\tt Student}$ and ${\tt TEMP3}$, i.e. the students enrolled 
with the lecturer attempting to execute the query. 
Therefore, the execution-time for
$\ulcorner{\tt SecQuery}({\cal S}, \texttt{Query\#6})\urcorner$ will increase 
depending on the number of students enrolled with the lecturer {\tt caller} and 
the ``complexity'' of the SQL expression 
${\rm map}({\rm auth}({\cal S}, r, {\rm read}({\tt Student:age})))$. 

Consider the case of the policy {\tt SecVGU\#2}.
Recall that the authorisation constraint ${\rm auth}(\texttt{SecVGU\#2}, {\tt Lecturer}, 
{\rm read}({\tt Enrolment}))$ is specified in OCL as follows:
\begin{tabbing}
${\tt (\mbox{\underline{\tt caller}} = \mbox{\underline{\tt lecturers}})\ or\ (\mbox{\underline{\tt caller}}.students\rightarrow{}{\tt exists}(s\mid{}s = \mbox{\underline{\tt students}}))}$.
 \end{tabbing}
   
Suppose that, as before,
we implement this authorisation constraint in SQL as follows:

\begin{lstlisting}[numbers=none]
(caller = lecturers)
OR (EXISTS (SELECT 1 FROM Enrolment e
           WHERE e.lecturers = caller
           AND e.students = students))
\end{lstlisting}

\noindent Notice then that, when executing
\begin{eqnarray*}
\begin{array}{l}
{\ulcorner {\rm SecQuery}(\texttt{SecVGU\#2}, \texttt{Query\#6})
\urcorner}({\tt caller}, \mbox{``{\tt Lecturer}''}),
\end{array}
\end{eqnarray*}
%
the SQL expression above will be executed once for
each row in the table {\tt TEMP2}, which is defined as the subset of the 
  cartesian product of the tables ${\tt Student}$ and ${\tt Lecturer}$
  that contains only the students of the lecturer attempting to
  execute the query, and that each time
this expression is executed, the clause 
\begin{lstlisting}[numbers=none]
WHERE e.lecturers = caller
AND e.students = students
\end{lstlisting}
will make a search among the rows in the table {\tt Enrolment}.

Moreover, recall that the authorisation constraint ${\rm auth}(\texttt{SecVGU\#2}, {\tt Lecturer},$ 
${\rm read}({\tt Student : age}))$ is specified in OCL as follows:
\begin{tabbing}
${\mbox{\underline{\tt caller}}{\tt .students}\rightarrow{}{\tt exists}{\tt (s\mid{}s =\ }\mbox{\underline{\tt self}})}$.
 \end{tabbing}
   
Suppose that this authorisation constraint
is implemented in SQL as follows:

\begin{lstlisting}[numbers=none]
EXISTS (SELECT 1 FROM Enrolment e
           WHERE e.lecturers = caller
           AND e.students = self)
\end{lstlisting}
%
%
\noindent Notice then that, when executing
\begin{eqnarray*}
\begin{array}{l}
{\ulcorner {\rm SecQuery}(\texttt{SecVGU\#2}, \texttt{Query\#6})
\urcorner}({\tt caller}, \mbox{``{\tt Lecturer}''}),
\end{array}
\end{eqnarray*}
%
the SQL expression above will be executed once for
each row in the table {\tt TEMP4}, which is defined as the join of the tables 
${\tt Student}$ and ${\tt TEMP3}$, i.e. the students enrolled 
with the lecturer 
attempting to execute the query, 
and that each time
this expression is executed, the clause 
\begin{lstlisting}[numbers=none]
WHERE e.lecturers = caller
AND e.students = self
\end{lstlisting}
will make a search among the rows in the table {\tt Enrolment}.
    
\paragraph{Possible optimisations} 

Suppose
that the user attempting to execute the query has the role ${\tt Lecturer}$.
In this case, the case-statement in lines 23--25 seems unnecessary, 
because:
\begin{itemize}
\item {\tt SecVGU\#2} authorises a lecturer to know his/her students,
\item the temporary table {\tt TEMP1} only contains  
students of the lecturer attempting to execute the query.
\end{itemize}
Applying the methodology described above, we can  prove that, 
in this case, the case-statement in lines 23--25 can be securely 
removed.
The SMT solver CVC4~\cite{DBLP:conf/cav/BarrettCDHJKRT11} solves this satisfiability problem 
in $0.057$ seconds.
The interested reader can find in Listing~\ref{app:cs3a:msfol} 
(Appendix~\ref{satisfiability:appendix})
the input to the CVC4 solver,
and in Listing~\ref{app:cs3:optsp} 
(Appendix~\ref{optimised:appendix})
the optimised version of the stored-procedure 
$\ulcorner{\rm SecQuery}({\cal S}, \texttt{Query\#6})\urcorner$.

Moreover, 
the case-statement in lines 38--39 also seems unnecessary, 
because:
\begin{itemize}
\item {\tt SecVGU\#2} authorises a lecturer to know the age of any student, if the student is his/her student, and
\item the temporary table ${\tt TEMP4}$ only contains  students of the 
lecturer attempting to execute the query.
\end{itemize}
Applying the methodology described above, and adding the fact
that the temporary table ${\tt TEMP4}$ only contains  students of the 
lecturer attempting to execute the query,
we can prove that, 
in this case, the case-statement in lines 38--39 can also be  securely 
removed.
%
 %
The SMT solver CVC4~\cite{DBLP:conf/cav/BarrettCDHJKRT11} solves this satisfiability problem 
in $0.038$ seconds.
The interested reader can find in Listing~\ref{app:cs3b:msfol}
(Appendix~\ref{satisfiability:appendix})
 the input to the CVC4 solver,
and in Listing~\ref{app:cs3:optsp} 
(Appendix~\ref{optimised:appendix})the optimised stored-procedure 
$\ulcorner{\rm SecQuery}({\cal S}, \texttt{Query\#6})\urcorner$.
%

Notice that the case-statements in lines 23--25 and 
38--39, cannot be removed, however, 
 for the case of the policies {\tt SecVGU\#1}.
 The interested reader can find in Listings~\ref{app:cs3a:msfol:x2} and~\ref{app:cs3b:msfol:x2} (both in Appendix~\ref{satisfiability:appendix}) 
 the 
satisfiability problems that correspond to these cases.

\section{Related work} \label{related-work:sec}
\label{related-work}
The work  presented here \emph{optimises} 
our  model-driven approach 
for enforcing FGAC policies when executing database queries~\cite{Bao2021}.
To the best of our knowledge, no directly related work exists yet.
Nevertheless, we discuss below  \emph{indirectly} related work:
namely,  proposals  related with our general approach for enforcing
FGAC policies.
To make this comparison concrete, we consider the implementation of our general approach in SQLSI.

The first  feature of our model-driven approach is that
it \emph{does not modify} the underlying database, except for adding the 
stored-procedures 
that configure our FGAC-enforcement mechanism.
This is in clear contrast with the
solutions  offered by the major commercial RDBMS,
which either recommend --- like in the case of MySQL or MariaDB~\cite{MariaDB10} ---
to manually create appropriate views
and  modify the queries so
as to referencing these views,
or they request --- like Oracle~\cite{Oracle02}, 
PostgreSQL~\cite{PostgreSQL}, and IBM~\cite{IBMDB2} --- 
to use  non-standard,  proprietary enforcement mechanisms.
As argued in~\cite{Bao2021}, 
the solutions currently offered 
by the major RDBMS are far from ideal: in fact, they are time-consuming,
error-prone, and scale poorly.

The second  feature of our model-driven approach 
is that
FGAC policies and  SQL queries are kept \emph{independent} 
of each other,
except for the fact that they refer to the same underlying
data model. This means, in particular, that FGAC policies can be
specified without knowing which SQL queries will be executed,
and vice versa. This is in clear contrast with
the solution recently proposed in~\cite{DBLP:conf/uss/MehtaEH0D17} where the
FGAC policies must be (re-)written depending on 
the  SQL queries that are executed. Nevertheless, 
the approach proposed in~\cite{DBLP:journals/jot/BaoC20}
 certainly shares with~\cite{DBLP:conf/uss/MehtaEH0D17}, as well as with other previous
approaches like~\cite{LeFevre04}, the idea
of enforcing FGAC-policies by \emph{rewriting} the SQL queries, 
instead of by modifying the underlying databases or
by using non-standard, proprietary  features.
 
The third  feature of our model-driven approach  is that
the enforcement mechanism can be \emph{automatically 
generated} from the FGAC-policies, by using available mappings
from OCL to SQL --- for example~\cite{BaoC19} ---
in order to implement the authorisation 
constraints appearing in the FGAC policies.
However, 
for the sake of execution-time performance,
we recommend
manually implementing in SQL the authorisation
constraints appearing in the FGAC policies.

\section{Conclusions and future work} \label{conclusions-future-work}

In~\cite{Bao2021} we  proposed  a model-driven approach
for enforcing fine-grained access control (FGAC) policies
when executing SQL queries.
  In a nutshell,  
  we defined a function ${\rm SecQuery}()$  that,
  given a  policy ${\cal S}$ and a query $q$, it
  generates a SQL stored-procedure,
  such that:
  if a user is authorised, according to ${\cal S}$,
  to execute $q$, then calling this stored-procedure
  will return the same result as executing $q$;
  otherwise, if a user is not authorised, according to ${\cal S}$,
  to execute $q$, then calling the stored-procedure
  will signal an error.
%

Since the stored-procedures generated by ${\rm SecQuery}()$
perform at execution-time the  authorisation checks required
by the given FGAC policy, not surprisingly, there is a significant loss in performance
when executing ``secured'' queries --- i.e., the stored-procedures
generated by ${\rm SecQuery}()$ --- with respect to executing  ``unsecured'' queries.
There are situations, however, in which performing some
authorisation checks may seem to be unnecessary.

In this article we have presented a general,
model-based approach that \emph{optimises} 
the ``secured'' queries generated
by ${\rm SecQuery}()$ by removing those
 authorisation checks that can be proved to be
 unnecessary. 
Moreover, we have  presented a concrete realisation 
of this  approach for our SQLSI
methodology for enforcing FGAC policies when
executing SQL queries. 
To prove in SQLSI that an authorisation check is
unnecessary, and therefore that it can be removed, 
we  formulate the corresponding problem 
as a satisfiability problem in many-sorted first-order logic,
and use SMT-solvers like CVC4~\cite{DBLP:conf/cav/BarrettCDHJKRT11} to try to solve it. 
To illustrate this approach
 we have  provided a non-trivial case study involving
 different FGAC policies, users, and queries.
  
  We recognise that the  SQLSI methodology needs
  to be  further developed, in several dimensions. 
 First of all, from the \emph{languages} point of view:
  we plan to extend our definition of data models 
  to include class generalisations; we also plan  to
  extend our definition of  FGAC security models
  to include role hierarchies and permissions for other types of actions,
  besides read actions;
 and we  plan to extend our definition of 
 ${\rm SecQuery}()$ to cover as much as possible of
 the SQL language, including, in particular, left/right-joins and
 group-by clauses.
 %
Secondly, from the \emph{code-generation} point of view, 
we plan to extend  SQLSI 
to cover also
insert, update, and delete statements.
 Thirdly, from the \emph{correctness} point of view, 
 we plan to develop a methodology for proving that OCL
 authorisation constraints are correctly implemented in SQL. 
  Finally, from the \emph{applicability} point of view, 
  we are interested in developing  a methodology \emph{\`a la}
 SQLSI for enforcing FGAC policies in
  the case of NoSQL databases.

\section*{Conflict of interest}
The authors declare that they have no conflict of interest.

\bibliographystyle{spmpsci}      
\bibliography{bib21}   

\appendix
\section{Case study. Satisfiability problems}
\label{satisfiability:appendix}

In this appendix we include the satisfiability problems 
discussed in our case study (Section~\ref{case-study}). 
Notice that these problems
refer to the same underlying  data model, namely,
the data model \texttt{University} (Figure~\ref{university:dm}). 
We show in Listing~\ref{dm:msfol}
below the MSFOL theory corresponding to the data model
{\tt University}.

\begin{lstlisting}[numbers = none,basicstyle=\ttfamily\small, label=dm:msfol, caption={{\tt University} data model theory},captionpos=b]
; sort declaration
(declare-sort Classifier 0)

; null and invalid object and its axiom
(declare-const nullClassifier Classifier)
(declare-const invalClassifier Classifier)
(assert (distinct nullClassifier invalClassifier))

; null and invalid integer and its axiom
(declare-const nullInt Int)
(declare-const invalInt Int)
(assert (distinct nullInt invalInt))

; null and invalid string and its axiom
(declare-const nullString String)
(declare-const invalString String)
(assert (distinct nullString invalString))

; unary predicate Lecturer(x) and its axiom
(declare-fun Lecturer (Classifier) Bool)
(assert (not (Lecturer nullClassifier)))
(assert (not (Lecturer invalClassifier)))

; unary predicate Student(x) and its axiom
(declare-fun Student (Classifier) Bool)
(assert (not (Student nullClassifier)))
(assert (not (Student invalClassifier)))

; axiom: disjoint set of objects of different classes
(assert (forall ((x Classifier)) 
    (=> (Lecturer x) (not (Student x)))))
(assert (forall ((x Classifier)) 
    (=> (Student x) (not (Lecturer x)))))

; function get the age of lecturer and its axiom
(declare-fun age_Lecturer (Classifier) Int)
(assert (= (age_Lecturer nullClassifier) invalInt))
(assert (= (age_Lecturer invalClassifier) invalInt))
(assert (forall ((x Classifier))
    (=> (Lecturer x)
        (distinct (age_Lecturer x) invalInt))))
	
; function get the email of lecturer and its axiom
(declare-fun email_Lecturer (Classifier) String)
(assert (= (email_Lecturer nullClassifier) invalString))
(assert (= (email_Lecturer invalClassifier) invalString))
(assert (forall ((x Classifier))
    (=> (Lecturer x)
        (distinct (email_Lecturer x) invalString))))
	
; function get the name of lecturer and its axiom	
(declare-fun name_Lecturer (Classifier) String)
(assert (= (name_Lecturer nullClassifier) invalString))
(assert (= (name_Lecturer invalClassifier) invalString))
(assert (forall ((x Classifier))
    (=> (Lecturer x)
        (distinct (name_Lecturer x) invalString))))
		
; function get the age of student and its axiom
(declare-fun age_Student (Classifier) Int)
(assert (= (age_Student nullClassifier) invalInt))
(assert (= (age_Student invalClassifier) invalInt))
(assert (forall ((x Classifier))
    (=> (Student x)
        (distinct (age_Student x) invalInt))))

; function get the name of student and its axiom		
(declare-fun name_Student (Classifier) String)
(assert (= (name_Student nullClassifier) invalString))
(assert (= (name_Student invalClassifier) invalString))
(assert (forall ((x Classifier))
    (=> (Student x)
        (distinct (name_Student x) invalString))))
		
; function get the email of student and its axiom
(declare-fun email_Student (Classifier) String)
(assert (= (email_Student nullClassifier) invalString))
(assert (= (email_Student invalClassifier) invalString))
(assert (forall ((x Classifier))
    (=> (Student x)
        (distinct (email_Student x) invalString))))
		
; binary predicate of the Enrolment association and its axiom
(declare-fun Enrolment (Classifier Classifier) Bool)
(assert (forall ((x Classifier))
    (forall ((y Classifier)) 
        (=> (Enrolment x y) 
            (and (Lecturer x) (Student y))))))
\end{lstlisting}

\subsection*{Case~\ref{cs1}} 

\begin{lstlisting}[numbers = none,basicstyle=\ttfamily\small, label=app:cs1:msfol, caption={Case~\ref{cs1}, {\tt SecVGU\#1}. The user is the oldest lecturer},captionpos=b, escapeinside={[*}{*]}]
; the generated theory for data model is exactly as in Listing[*~\ref{dm:msfol}*]

; constant symbol of caller and its axiom
(declare-const caller Classifier)
(assert (Lecturer caller))

; constant symbol of self and its axiom
(declare-const self Classifier)
(assert (Student self))

; caller property: caller is indeed the oldest lecturer
; Lecturer.allInstances()->forAll(l|l.age <= caller.age)
(assert (forall ((l Classifier)) 
    (and (=> (Lecturer l) 
             (and (<= (age_Lecturer l) (age_Lecturer caller)) 
                  (not (or (= (age_Lecturer l) nullInt) 
                           (or (= l nullClassifier) 
                               (= l invalidClassifier)) 
                           (= (age_Lecturer caller) nullInt) 
                           (or (= caller nullClassifier) 
                               (= caller invalidClassifier)))))) 
         (not false))))

; this TEMP0 function is the OCL expression
; Lecturer.allInstances()->select(l|l.age > caller.age)
(declare-fun TEMP0 (Classifier) Bool)
(assert (forall ((l Classifier)) 
  (= (TEMP0 l) 
     (and (Lecturer l) 
          (and (> (age_Lecturer l) (age_Lecturer caller)) 
               (not (or (= (age_Lecturer l) nullInt) 
                        (or (= l nullClassifier) 
                            (= l invalidClassifier)) 
                        (= (age_Lecturer caller) nullInt) 
                        (or (= caller nullClassifier) 
                            (= caller invalidClassifier)))))))))

; authorisation constraint [*${\it auth}$*]: caller is the oldest lecturer
; Lecturer.allInstances()->select(l|l.age > caller.age)->isEmpty()
; below is the negation of [*${\rm map}_{\rm true}({\it auth})$*]
(assert (not (forall ((x Classifier)) 
                    (and (not (TEMP0 x)) (not false)))))
\end{lstlisting}

\begin{lstlisting}[numbers = none,basicstyle=\ttfamily\small, label=app:cs1:msfol:x3, caption={Case~\ref{cs1}, {\tt SecVGU\#2}. The user is the oldest lecturer},captionpos=b, escapeinside={[*}{*]}]
; the generated theory for data model is exactly as in Listing[*~\ref{dm:msfol}*]

; constant symbols of caller, self and its axiom
; are  defined as in Listing[*~\ref{app:cs1:msfol}*]

; caller property: caller is indeed the oldest lecturer
; is  defined as in Listing[*~\ref{app:cs1:msfol}*]

; authorisation constraint [*${\it auth}$*]: a caller can know the age of any student
; if the caller is the lecturer of that student
; caller.students->exists(s|s = students)
; below is the negation of [*${\rm map}_{\rm true}({\it auth})$*]
(assert (not (exists ((temp Classifier)) 
    (and (Enrolment caller temp) 
         (= temp self) 
         (not (or (= caller nullClassifier) 
                  (= caller invalidClassifier))) 
         (not (= self invalidClassifier))))))
\end{lstlisting}

\subsection*{Case~\ref{cs2}}


\begin{lstlisting}[numbers = none,basicstyle=\ttfamily\small, label=app:cs2:msfol, caption={Case~\ref{cs2}, {\tt SecVGU\#2}.
The user is the lecturer of every student},captionpos=b, escapeinside={[*}{*]}]
; the generated theory for data model is exactly as in Listing[*~\ref{dm:msfol}*]

; constant symbol of caller and its axiom
(declare-const caller Classifier)
(assert (Lecturer caller))

; constant symbol of lecturers and its axiom
(declare-const lecturers Classifier)
(assert (Lecturer lecturers))

; constant symbol of students and its axiom
(declare-const students Classifier)
(assert (Student students))

; caller property: caller is the lecturer of every student
; Student.allInstances()->forAll(s|s.lecturers->includes(caller)
(assert (forall ((s Classifier)) 
    (and (=> (Student s) 
             (exists ((temp Classifier)) 
                 (and (Enrolment temp s) 
                      (= temp caller) 
                      (not (or (= s nullClassifier) 
                               (= s invalidClassifier))) 
                      (not (= caller invalidClassifier))))) 
         (not false))))

; authorisation constraint [*${\it auth}$*]: a lecturer can know his/her student and
; can know the students of any lecturer if the student is his/her student
; caller = lecturers or caller.students->includes(students)
; below is the negation of [*${\rm map}_{\rm true}({\it auth})$*]
(assert (not (or (or (and (= caller nullClassifier) 
                          (= lecturers nullClassifier)) 
                     (and (= caller lecturers) 
                          (not (or (= caller nullClassifier) 
                                   (= caller invalidClassifier) 
                                   (= lecturers nullClassifier) 
                                   (= lecturers invalidClassifier))))) 
                 (exists ((temp Classifier)) 
                     (and (Enrolment temp students) 
                          (= temp caller) 
                          (not (or (= students nullClassifier) 
                                   (= students invalidClassifier))) 
                          (not (= caller invalidClassifier)))))))
\end{lstlisting}


\begin{lstlisting}[numbers = none,basicstyle=\ttfamily\small, label=app:cs2:msfol:x2, caption={Case~\ref{cs2}, {\tt SecVGU\#1}. The user is the lecturer of every student},captionpos=b, escapeinside={[*}{*]}]
; the generated theory for data model is exactly as in Listing[*~\ref{dm:msfol}*]

; constant symbol of caller, lecturers, students and its axiom
; are defined as in Listing[*~\ref{app:cs2:msfol}*]

; caller property: caller is indeed the oldest lecturer
; is defined as in Listing[*~\ref{app:cs2:msfol}*]

; this TEMP0 function is the OCL expression
; Lecturer.allInstances()->select(l|l.age > caller.age)
; is  defined as in Listing[*~\ref{app:cs1:msfol}*]

; authorisation constraint [*${\it auth}$*]: caller is the oldest lecturer
; Lecturer.allInstances()->select(l|l.age > caller.age)->isEmpty()
; the negation of this [*${\it auth}$*] is  defined as in Listing[*~\ref{app:cs1:msfol}*]
\end{lstlisting}

\subsection*{Case~\ref{cs3}}


\begin{lstlisting}[numbers = none,basicstyle=\ttfamily\small, label=app:cs3a:msfol, caption={Case~\ref{cs3}~(I), {\tt SecVGU\#2}. 
The user has  role {\tt Lecturer}},captionpos=b, escapeinside={[*}{*]}]
; the generated theory for data model is exactly as in Listing[*~\ref{dm:msfol}*]

; constant symbol of caller, lecturers, students and its axiom
; are  defined as in Listing[*~\ref{app:cs2:msfol}*]

; caller is the lecturer in the considered records
; caller = lecturers
(assert (or (and (= caller nullClassifier) 
                 (= lecturers nullClassifier)) 
            (and (= caller lecturers) 
                 (not (or (= caller nullClassifier) 
                          (= caller invalidClassifier) 
                          (= lecturers nullClassifier) 
                          (= lecturers invalidClassifier))))))

; authorisation constraint [*${\it auth}$*]: a lecturer can know his/her student and
; can know the students of any lecturer if the student is his/her student
; caller = lecturers or caller.students->includes(students)
; the negation of this auth is  defined as in Listing[*~\ref{app:cs2:msfol}*]
\end{lstlisting}


\begin{lstlisting}[numbers = none,basicstyle=\ttfamily\small, label=app:cs3a:msfol:x2, caption={Case~\ref{cs3}~(I), {\tt SecVGU\#1}. 
The user has  role {\tt Lecturer}},captionpos=b, escapeinside={[*}{*]}]
; the generated theory for data model is exactly as in Listing[*~\ref{dm:msfol}*]

; constant symbol of caller, lecturers, students and its axiom
; are  defined as in Listing[*~\ref{app:cs2:msfol}*]

; caller is the lecturer
; OCL: caller = lecturers
; is  defined as in Listing[*~\ref{app:cs3a:msfol}*]

; this TEMP0 function is the OCL expression
; Lecturer.allInstances()->select(l|l.age > caller.age)
; is  defined as in Listing[*~\ref{app:cs1:msfol}*]

; authorisation constraint [*${\it auth}$*]: caller is the oldest lecturer
; Lecturer.allInstances()->select(l|l.age > caller.age)->isEmpty()
; the negation of this [*${\it auth}$*] is  defined as in Listing[*~\ref{app:cs1:msfol}*]
\end{lstlisting}

\begin{lstlisting}[numbers = none,basicstyle=\ttfamily\small, label=app:cs3b:msfol, caption={Case~\ref{cs3}~(II), {\tt SecVGU\#2}.
The user has  role {\tt Lecturer}},captionpos=b, escapeinside={[*}{*]}]
; the generated theory for data model is exactly as in Listing[*~\ref{dm:msfol}*]

; constant symbols of caller, self and its axiom
; are  defined as in Listing[*~\ref{app:cs1:msfol}*]

; the students considered are students of the caller
; caller.students->includes(self)
(assert (exists ((temp Classifier)) 
    (and (Enrolment caller temp) 
         (= temp self) 
         (not (or (= caller nullClassifier) 
                  (= caller invalidClassifier))) 
         (not (= self invalidClassifier)))))

; authorisation constraint [*${\it auth}$*]: a caller can know the
; age of any student, if the caller is the lecturer or of that student
; caller.students->exists(s | s = self)
; below is the negation of [*${\rm map}_{\rm true}({\it auth})$*]
(assert (not (exists ((temp Classifier)) 
    (and (Enrolment caller temp) 
         (= temp self) 
         (not (or (= caller nullClassifier) 
                  (= caller invalidClassifier))) 
         (not (= self invalidClassifier))))))
\end{lstlisting}

\begin{lstlisting}[numbers = none,basicstyle=\ttfamily\small, label=app:cs3b:msfol:x2, caption={Case~\ref{cs3}~(II), {\tt SecVGU\#1}.
The user has  role {\tt Lecturer}},captionpos=b, escapeinside={[*}{*]}]
; the generated theory for data model is exactly as in Listing[*~\ref{dm:msfol}*]

; constant symbols of caller, self and its axiom
; are  defined as in Listing[*~\ref{app:cs1:msfol}*]

; the students considered are students of the caller
; OCL: caller.students->includes(self)
; is  defined as in Listing[*~\ref{app:cs3b:msfol}*]

; this TEMP0 function is the OCL expression
; Lecturer.allInstances()->select(l|l.age > caller.age)
; is  defined as in Listing[*~\ref{app:cs1:msfol}*]

; authorisation constraint [*${\it auth}$*]: caller is the oldest lecturer
; Lecturer.allInstances()->select(l|l.age > caller.age)->isEmpty()
; the negation of this [*${\it auth}$*] is  defined as in Listing[*~\ref{app:cs1:msfol}*]
\end{lstlisting}

\section{Optimised stored-procedures}
\label{optimised:appendix}
\subsection*{Case~\ref{cs1}} 
We can  enforce the policy {\tt SecVGU\#1} by 
using the following if-then-else (Listing~\ref{app:cs1:optsp}):
if the user is the oldest lecturer, then  we execute the original
query {\tt Query\#4}, without further checks; 
otherwise, we execute the ``securized'' query corresponding 
to {\tt Query\#4}.
\begin{lstlisting}[numbers = none,basicstyle=\ttfamily\small, label=app:cs1:optsp,  caption={Case~\ref{cs1}, {\tt SecVGU\#1}},captionpos=b, escapeinside={[*}{*]}]
% declare and assign the variable caller.
% declare and assign the variable role.
IF (role = 'Lecturer'
    AND ((SELECT MAX(age) FROM Lecturer)
            = (SELECT age FROM Lecturer WHERE Lecturer_id = caller)))
THEN
  % if the condition is satisfied, i.e. caller is the oldest lecturers,
  % then the case-statement is removed.
  CREATE TEMPORARY TABLE TEMP1 AS (
    SELECT * FROM Student WHERE age > 18
  );
ELSE
  % otherwise, the case-statement as before.
  CREATE TEMPORARY TABLE TEMP1 AS (
    SELECT * FROM Student
    WHERE CASE [*$\ulcorner{\rm AuthFunc}({\cal S}, {\tt Student:age})\urcorner$*](caller, role, Student_id)
      WHEN 1 THEN age ELSE throw_error() END > 18
  );
END IF;
\end{lstlisting}

\subsection*{Case~\ref{cs2}}
We can enforce the policy {\tt SecVGU\#2} by 
using the following if-then-else (Listing~\ref{app:cs2:optsp}): 
if the user is a lecturer of every student, 
then we execute the original
query {\tt Query\#5}, without further checks; 
otherwise, we execute the ``securized'' query 
corresponding to {\tt Query\#5}.
\begin{lstlisting}[numbers = none,basicstyle=\ttfamily\small, label=app:cs2:optsp, caption={Case~\ref{cs2}, {\tt SecVGU\#2}},captionpos=b, escapeinside={[*}{*]}]
% declare and assign the variable caller.
% declare and assign the variable role.
IF (role = 'Lecturer'
    AND ((SELECT COUNT(*) FROM Student)
            = (SELECT COUNT(*) FROM Enrolment WHERE lecturers = caller)))
THEN
  % if the condition is satisfied, 
  % i.e. caller is the lecturer of every student,
  % then the case-statement is removed.
  CREATE TEMPORARY TABLE TEMP2 AS (
    SELECT * FROM TEMP1 WHERE TRUE
  );
ELSE
  % otherwise, then the case-statement as before.
  CREATE TEMPORARY TABLE TEMP2 AS (
    SELECT * FROM TEMP1
    WHERE CASE [*$\ulcorner{\rm AuthFunc}({\cal S},$ ${\tt Enrolment})\urcorner$*](caller, role, lecturers,
      students) WHEN TRUE THEN TRUE ELSE throw_error() END
  );
END IF;
\end{lstlisting}

\subsection*{Case~\ref{cs3}}
We can enforce  the policy 
${\tt SecVGU\#2}$ by 
using the  if-then-else statements shown in
Listing~\ref{app:cs3:optsp}.

\begin{lstlisting}[numbers = none,basicstyle=\ttfamily\small, label=app:cs3:optsp, caption={Case~\ref{cs3}, {\tt SecVGU\#2}},captionpos=b, escapeinside={[*}{*]}]
% declare and assign the variable caller.
% declare and assign the variable role.
IF (role = 'Lecturer')
THEN
  % if the condition is satisfied, 
  % i.e. caller has role Lecturer,
  % then the case-statement is removed.
  CREATE TEMPORARY TABLE TEMP2 AS (
    SELECT * FROM TEMP1 WHERE TRUE
  );
ELSE
  % otherwise, then the case-statement as before.
  CREATE TEMPORARY TABLE TEMP2 AS (
    SELECT * FROM TEMP1
    WHERE CASE [*$\ulcorner{\rm AuthFunc}({\cal S},$ ${\tt Enrolment})\urcorner$*](caller, role, lecturers,
      students) WHEN TRUE THEN TRUE ELSE throw_error() END
  );
END IF;
...
IF (role = 'Lecturer')
THEN
  % if the condition is satisfied, 
  % i.e. caller has role Lecturer,
  % then the case-statement is removed.
  CREATE TEMPORARY TABLE TEMP5 AS (
    SELECT age FROM TEMP4
  );

ELSE
  % otherwise, then the case-statement as before.
  CREATE TEMPORARY TABLE TEMP5 AS (
    SELECT CASE [*$\ulcorner{\rm AuthFunc}({\cal S},$ ${\tt Student:age})\urcorner$*](caller, role, Student_id)
      WHEN 1 THEN age ELSE throw_error() END as age 
    FROM TEMP4
  );
END IF;
\end{lstlisting}

\end{document}